\documentclass{pazh_engl}

\usepackage{epsf}
\usepackage{flushrt}
\usepackage{graphicx}
\usepackage{pstricks}

\begin{document}

\sloppypar

\title{\bf RXTE observations of strongly absorbed sources IGR J16318-4848 and 
IGR J16358-4726}

\author{\copyright 2003 Ç. M. Revnivtsev\inst{1,2}}

\institute{Max-Planck-Institut fuer Astrophysik, Garching, Germany
\and
Space Research Institute, Moscow, Russia}

\authorrunning{REVNIVTSEV}
\titlerunning{}
\date{April 18, 2003}
\abstract{Results of analysis of RXTE observations of strongly absorbed 
X-ray sources IGR J16318-4848 and IGR J16358-4726 are presented.
Careful subtraction of Galactic ridge emission contribution to the observed
spectra of RXTE/PCA allowed us to obtain the spectra of the sources in 3--25 keV 
energy band. Spectra of the sources cold be well described by a power law
with photoabsorption. It is shown that the value of absorption
column in the case of RXTE observation of IGR J16318-4848 performed on
March 14.1, 2003 is somewhat higher  that that obtained by XMM observatory
on Feb 10.7 2003. This could imply that the source has variable absorption,
presumably connected with an orbital phase of the binary system. It is noted,
that all three X-ray sources, discovered by INTEGRAL observatory
in the sky region of $(l,b)\sim(336,0)$ (IGR J16318-4848, IGR 
J16320-4751 and IGR J16358-4726) have large intrinsic photoabsorption and
could be high mass X-ray binaries. This hypothesis has indirect support
from the fact that their are located close to the Norma spiral arm tangent, 
i.e. in the region of enhanced concentrations of young massive stars. If they
are reside within this spiral arm some rough estimation of the sources 
distance could be made -- $D\sim$6-8 kpc.
}
\titlerunning{RXTE observations of IGR J16318-4848 and IGR J16358-4726}
\maketitle

\section*{Introduction}
During first several months of Galactic plane scans of INTEGRAL observatory
there were discovered a few sources which X-ray spectrum has high
photoabsorption -  IGR J16318-4848 (Courvoisier et al. 2003, Murakami et al. 2003),
IGR J16320-4751/AX J1631.9-4752 (Tomsick et al. 2003), IGR J16358-4726 
(Revnivtsev et al. 2003a,b)

Shortly after disovery of IGR J16318-4848, the source was observed by XMM
 (Schartel et al. 2003, de Plaa et al. 2003), that allowed  one to make
precise position determination and identify its ifrared and optical 
companion (Foschini et al. 2003, Revnivtsev et al. 2003c).
Analysis of infrared and optical measurements of the counterpart showed that
X-ray binary IGR J16318-4848 likely contains supergiant companion, which
powerful stellar wind could lead to the observed properties of the source
 - very strong photoabsorption and very powerful fluorescent emission 
lines of neutral iron - very similar to that of long period pulsar
GX301-2 (Revnivtsev et al. 2003c).

Observations of IGR J16320-4751 and IGR J16358-4726 with XMM and CHANDRA 
satellites (Rodriguez et al. 2003, Kouveliotou et al. 2003) have not 
resulted in unambiguous identification of their infrared counterparts, 
however, in both cases companion stars are likely to be bright, and
possibly massive.

In the present Letter we present results of analysis of RXTE observations
of IGR J16318-4848 and IGR J16358-4726.

\section*{Data analysis and results}

IGR J16318-4848 and IGR J16358-4726 were observed by RXTE on March 14.1, 2003
 and March 25.9, 2003 respectively. Effective exposure times of these 
observations were 6.5 ksec and 3.1 ksec. Because sources are located in 
highly populated area of the sky and the field of view of RXTE/PCA 
instrument is 1$^\circ$, the orientation of the observatory in both cases
had some offset with respect to the real source positions, in order to
exclude possible contribution of nearby known bright sources. Map 
of the sky around sources with overplotted fields of view of RXTE/PCA
during two observations is presented in Fig.1.

\begin{figure}
\includegraphics[width=\columnwidth]{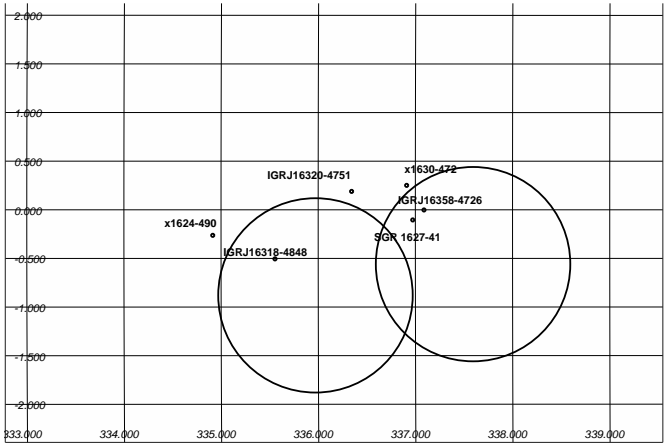}
\caption{Map of the X-ray sky around IGR J16318-4848 
and IGR J16358-4726. Big circles denote fields of view of RXTE/PCA during 
two analized observations.}
\end{figure}

For the data reduction we used standard programs of package FTOOLS/LHEASOFT
 5.2. Because the sources of our interest are rather weak and they were 
observed with offset pointings, that diminshes the flux detected  
from them, we have paid special attention to the correct PCA background
subtraction. First of all we have excluded the data from PCU0, because
from May 2000 this detector lacks propane veto layer, that leads to
the worse quality of background subtraction. Besides, we have used data from
upper layer of anodes (LR1) only, because it has lower systematic 
uncertainties of the background estimation. 
For the background modelling we used model ''L7\_240''.

Galactic Ridge emission very strongly contributes to the X-ray 
emission detected by RXTE/PCA, that complicates the analysis of 
weak sources close to the Galactic plane. For example, the X-ray flux, detected by PCA from 
IGR J16318-4848 in 2-10 keV energy range is not more than 0.1-0.2 mCrab
however Galactic ridge emission, integrated over PCA field of view,
gives X-ray flux at the level of approximately 3--5 mCrab (Valinia, Marshall
 1998, Revnivtsev 2003).

In order to take into account the Galactic ridge emission we have
used results of analysis, presented in papers of Valinia, Marshall (1998)
and Revnivtsev (2003). Besides, in our analysis we used RXTE/PCA data 
obtained during observation of the point, close to IGR J16318-4848 
and  IGR J16358-4726, and namely  - observations of SGR 1627-41 in a 
quiescent state (Nov. 19-20, 2001). As the shape of the spectrum of 
Galactic ridge emission does not depend on the position on the sky,
except for changing value of interstellar photoabsorption (Yamasaki et 
al. 1997, Valinia, 
Marshall 1998, Tanaka 2002, Revnivtsev 2003), we can use its spectrum 
obtained in an empty field as a template in order to subtract its
contribution to the observations of IGR J16318-4848 and IGR J16358-4726.

\subsection*{IGR J16318-4848}

Spectrum of IGR J16318-4848, averaged over whole observation is presented
in Fig.2. In the figure we also present total spectrum, detected by 
RXTE/PCA, that consists of the spectrum of the source and the spectrum
of Galactic ridge emission. Note, that normalization of the source 
spectrum is in  $\sim2.22$ times smaller than the real one because of
$\sim0.55^\circ$ offset pointing of RXTE/PCA.

For the source spectral approximation we have used simplest model -- 
a power law ($dN(E)\propto E^{-\alpha}dE$) with neutral 
photoabsorption ($wabs$ model is XSPEC package). This model
was successfully used for spectral approximation of XMM and ASCA data
(de Plaa et al. 2003, Revnivtsev et al. 2003c). It is worth to note, 
that in spite of known presence of prominent fluorescent emission lines
at 6--7 keV (Schartel et al. 2003, de Plaa et al. 2003, Revnivtsev et al. 
2003c) very strong influence of Galactic ridge emission at these
energies precludes any analysis of these lines in RXTE/PCA data.
Flux from the Galactic ridge emission, detected by RXTE/PCA at energies 
6--7 keV is more than 10 times higher than that of IGR J16318-4848 
(see. Fig.2). Obtained best fit parameters of the source spectrum
are presented in Table 1. Note, that the parameter of neutral photoabsorption
$n_HL$ is somewhat larger than that obtained by XMM on Feb 10.7, 2003
 (de Plaa et al. 2003, Matt, Guainazzi 2003).

\begin{figure}
\includegraphics[width=\columnwidth]{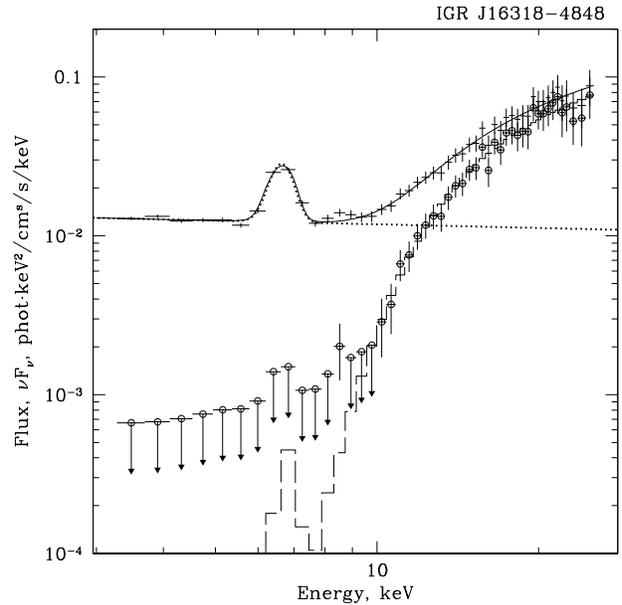}
\caption{Spectrum of IGR J16318-4848. Crosses denote spectrum detected by
RXTE/PCA, open circles - extracted spectrum of IGR J16318-4848 after
subtraction of Galactic ridge emission. Solid curve represents the model
of RXTE/PCA observed spectrum, that consists of model of Galactic ridge 
emission (dotted curve) and model of the source spectrum (dashed curve).}
\end{figure}

The lightcurve of the source over the whole obsevation is presented
in Fig.3. 

Analysis of the power spectrum of the obtained lightcurve have not
shown any pulsations or highly coherent oscillations (Swank, Markwardt 2003). 
2 $\sigma$ upper
limit on the amplitude of possible pulsations with a frequencies
0.01 Hz - 1 kHz in the energy band 10-20 keV is approximately 10-15\%.
Upper limit on an amplitude of pulsations with longer periods, 
like in the case of GX301-2 (670 sec) or IGR J16358-4726 (5860 sec, 
Kouveliotou et al. 2003) is even larger, $\sim$20-50\%. It is not
possible to study the light curve of the source at  energies lower than
$\sim$10 keV because of the source weakness with respect to the 
Galactic ridge emission.

\begin{figure}
\includegraphics[width=\columnwidth]{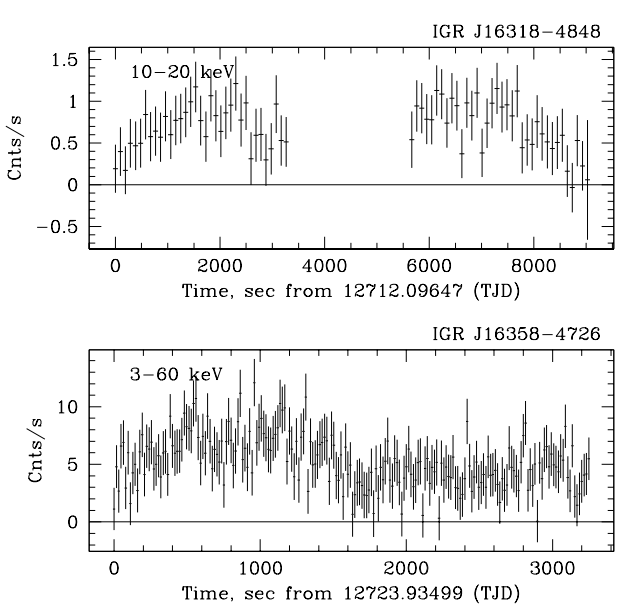}
\caption{Lightcurve of IGR J16318-4848 and IGR J16358-4726. Contribution 
of the galactic ridge emission is subtracted.}
\end{figure}

\begin{table*}
\caption{Best fit parameters of the spectra of IGR J16318-4848 and 
IGR J16358-4726, obtained with RXTE/PCA}
\tabcolsep=9mm
\begin{tabular}{l|c|c}
Parameter&IGR J16318-4848&IGR J16358-4726\\
\hline
Absorption column $N_H L$, $10^{22}$ cm$^{-2}$&$310\pm70$&$40\pm10$\\
Photon index $\alpha$&$1.0\pm0.5$&$1.1\pm0.2$\\
Observed flux$^a$ (3-25 keV), $10^{-10}$erg/s/cm$^2$&1.1&5.4\\ 
\hline
\end{tabular}
\begin{list}{}
\item $^a$ Observed flux of the source was corrected for the collimator responce
\end{list}
\end{table*}

\subsection*{IGR J16358-4726}

Spectrum of the source, averaged over whole observation is presented in Fig. 4.
Similar to the case of IGR J16318-4848 on the Fig.4 we present total
spectrum, detected by RXTE/PCA, that consist of the source spectrum and
the spectrum of Galactic ridge emission.

For the spectral approximation we used the same model as for the case of
IGR J16318-4848 - power law with neutral photoabsorption. Best fit
parameters of used model are presented in Table 1. Obtained values 
of parameters well agree with CHANDRA results (Kouveliotou et al. 2003)

As it was expected, the flux of the Galactic ridge emission in this
observation is higher than in the case of IGR J16318-4848 (Fig.2,4)
because the field of view of RXTE/PCA was $\sim$0.5$^\circ$ closer 
to the Galactic plane (see Fig.1). The value of the flux of the Galactic 
ridge emission well agrees with the results of Valinia, Marshall (1998), 
Revnivtsev (2003).

\begin{figure}
\includegraphics[width=\columnwidth]{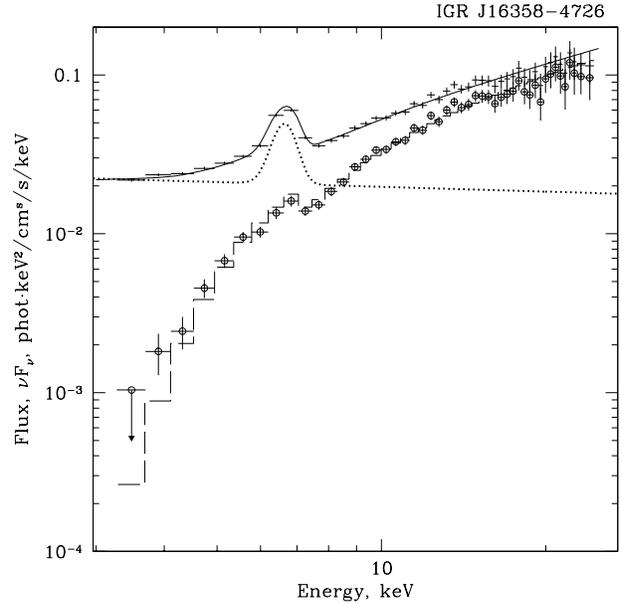}
\caption{Spectrum of IGR J16358-4726. Crosses denotes the spectrum, observed
by RXTE/PCA, open circles - spectrum of IGR J16358-4726 after subtraction 
of Galactic ridge emission. Solid curve represents the model
of RXTE/PCA observed spectrum, that consists of model of Galactic ridge 
emission (dotted curve) and model of the source spectrum (dashed curve).}
\end{figure}

Lightcurve of the source does not allow us to analize pulsations with
the period 5.86 ksec, detected by Kouveliotou et al. (2003), because
the length of our observation is only 3.1 ksec. Pulsations or highly coherent
oscillations with shorter periods was not detected. 2$\sigma$ upper limit
on such oscillations at frequencies 0.01Hz -- 1 kHz is $\sim$10-15\%.

\section*{Discussion}

In the previous chapter we presented results of analysis of RXTE observations
of two strongly
absorbed X-ray sources IGR J16318-4848 and IGR J16358-4726.
In spite of faitness of sources, strong influence of the Galactic ridge
emission and closeness of other bright X-ray binaries, the chosen
observation strategy allowed us to obtain the spectra of sources in 
3--25 keV energy band.

Spectra of considered sources could be well described by a power law 
with neutral photoabsorption (model $wabs * power$ of XSPEC package).
Because of large contribution of Galactic ridge emission (that 
contains a set of powerful lines in 6-7 keV energy band) 
to the spectra detected by RXTE/PCA, it is not possible to study
the emission lines, that could be present in the spectra of
IGR J16318-4848 and IGR J16358-4726. Obtained best fit parameters of our 
spectral approximation well agree with results of other observatories, except 
for somewhat higher value of absorption column $n_HL$ in the case of  IGR J16318-4848
with respect to the value obtained with XMM observatory (de Plaa et al. 
2003, Matt, Guainazzi 2003).

In spite of considerable complications that arises from strong contribution
 of Galactic ridge emission to the observed spectra of RXTE/PCA we
still believe that our obtained best fit parameters of spectral
approximations are more or less correct. Compatibility of best fit 
parameters of the spectrum of IGR J16358-4726 obtained by RXTE/PCA on March 
25.9, 2003 (see Table 1) and obtained by CHANDRA on March 24.2, 2003 
(Kouveliotou et  al. 2003) could serve as a demontration of correctness 
of our method of spectral analysis.

Therefore we suppose that higher value (with respect to the value obtained 
by XMM observatory on Feb. 10.7, 2003) of obtained absorption column
in the case of IGR J16318-4848 could imply that the source has variable 
absorption, probably connected with the orbital phase of the binary system.
Simiar absorption variability was observed for a large set of high mass X-ray 
binary systems, in particular in the case of long period pulsar GX 301-2 
(see e.g. Endo et al. 2002) which observational appearances are very similar 
to that of IGR J16318-4848.

It is  interesting to note, that three X-ray sources, discovered by INTEGRAL
observatory in the region of $(l,b)\sim(336,0)$ (IGR J16318-4848, 
IGR J16320-4751 and IGR J16358-4726) have a lot of things in common.
All three sources have strongly absorbed X-ray spectra, values of the 
absorption column strongly exceed the interstellar ones (e.g. Dickey, 
Lockman 1990). All three sources have rather hard X-ray spectra 
(photon index $\alpha\sim1-1.3$), that is typical for X-ray pulsars and
high mass binaries. For IGR J16318-4848 it was shown that
its optical/infrarred counterpart is bright, likely massive, star
(Foschini et al. 2003, Revnivtsev et al. 2003c). For other two sources
no unambiguous identification of counterparts were made yet, however it is 
likely that these sources also have bright companions, similar to
 IGR J16318-4848 (see. Rodriguez et al. 2003, Kouveliotou et al. 2003).
Therefore it seems to be reasonable to assume that all three X-ary sources
could be high miass binaries. Location of the sources on the sky could 
also be treated in the favor of described hypothesis. Sources are located 
at $(l,b)\sim(336,0)$, close to the Norma spiral arm tangent, i.e.
in the region of enhanced concentration of young massive stars (see e.g.
Grimm et al. 2002). If our assumption is correct than the rough estimation 
of the sources distance could be made - 6--8 kpc.

\bigskip
{\it

Author is very grateful to Dr. Jean Swank and RXTE planning team for RXTE
 TOO observations of IGR J16318-4848 and IGR J16358-4726; to Marat
Gilfanov for usefull discussions. Research has made use of data obtained 
through the High Energy Astrophysics Science Archive Research Center 
Online Service, provided by the NASA/Goddard Space Flight Center.
}
\section*{REFERENCES}

\indent
Courvoisier T., Walter R., Rodriguez J., Bouchet L.,
Lutovinov A., IAU Circ. 8063 (2003) 

Dickey J.M., Lockman F.J., 
Ann. Rev. Astron. Astrophys. {\bf 28}, 215 (1990)

Endo T., Ishida M., Masai K. et al., Astrophys. J. {\bf
574}, 897 (2002)

Foschini L., Rodriguez J., Walter R., IAU Circ. 8076 (2003)

Grimm H.-J., Gilfanov M., Sunyaev R., Astron.Astroph {\bf 391}, 923 (2002)

Kouveliotou C., Patel S., Tennan A. et al., IAUC 8109 (2003)

Matt G., Guainazzi M.,  MNRAS, accepted (2003), astro-ph/030362

Murakami H., Dotani T., Wijnands R.,  IAU Circ. 8070 (2003)

de Plaa J., den Hartog P., Kaastra J., in 't Zand J.,
Mendez M., Hermsen W.,  Astronomer's Telegram 119 (2003) 

Revnivtsev 2003, A\&A sumbmitted, astro-ph/0304351

Revnivtsev M., Tuerler M., Del Santo M. et al.,  IAUC 8097 (2002a) 

Revnivtsev M., Lutovinov A., Ebisawa K.,  Astronomer's 
Telegram 131 (2003b)

Revnivtsev M., Sazonov S., Gilfanov M., Sunyaev R., Astronomy Letters, 
accepted (2003c), astro-ph/0303274

Rodiriguez J., Tomsick J., Foschini L. et al., A\&A submitted, astro-ph/0304139

Schartel N., Ehle M., Breitfellner M. et al., IAU
Circ. 8072 (2003)

Swank J., Markwardt C., Astronomer's Telegram 128 (2003)

Tanaka Y., Astron.Astroph, {\bf 382}, 1052 (2002)

Valinia A., Marshall F., Astroph. J. {\bf 505}, 134 (1998)

Yamasaki N., Ohashi T., Takahara F. et al., Astroph.J, {\bf 481}, 821 (1997)

\end{document}